\documentstyle[prd,aps,psfig]{revtex}

\begin{document}

\draft

\title{On the Numerical Stability of the Einstein Equations}

\author{
Mark~Miller \medskip
}

\address{
McDonnell Center for the Space Sciences \\
Department of Physics,
Washington University, St. Louis, Missouri 63130}

\date{\today}

\maketitle

\begin{abstract}
 
  We perform a von Neumann stability analysis on a common discretization
  of the Einstein equations.  The analysis is performed on two formulations
  of the Einstein equations, namely, the standard ADM formulation
  and the conformal-traceless (CT) formulation.  The eigenvalues
  of the amplification matrix are computed for flat space 
  as well as for a highly nonlinear plane wave exact solution.  We find 
  that for the flat space initial data, the condition for stability
  is simply $\frac {\Delta t}{\Delta z} \leq 1$.
  However, a von Neumann analysis for highly nonlinear plane wave 
  initial data shows that the standard
  ADM formulation is unconditionally unstable, while the 
  conformal-traceless (CT) formulation is stable for 
  $0.25 \leq \frac {\Delta t}{\Delta z} < 1$.

\end{abstract}


%\vskip2pc]

%\narrowtext

%%%%%%%%%%%%%%%%%%%%%%%%
%%%   INTRODUCTION   %%%
%%%%%%%%%%%%%%%%%%%%%%%%

\section{Introduction}
\label{sec:Introduction}

Numerically solving the full 3D nonlinear Einstein equations is, for
several reasons, a daunting task.  Still, numerical relativity 
remains the best method for studying astrophysically interesting regions 
of the solution space of the 
Einstein equations in sufficient detail and accuracy in order 
to be used to interpret measurements made by the up and coming
gravitational wave detectors.  Even though numerical relativity
is almost 35 years old, some of the same problems faced by
researchers three decades ago are present today.  Aside from 
the computational complexity of implementing a numerical solver for 
the nonlinear Einstein equations, there exist several unsolved 
problems, including the well-posedness of certain initial value 
formulations of the Einstein equations and the proper choice of
gauge.  Not the least of these problems is numerical stability.  
A common thread in numerical relativity research over the past
three decades is the observation of high frequency (Nyquist 
frequency) noise growing and dominating the numerical solution.  
Traditionally, numerical studies have been performed with the initial
value formulation of the Einstein equations known as the ADM 3+1
formulation\cite{Arnowitt62}, 
in which the 3-metric and extrinsic curvature are
the dynamically evolved variables.  

Lately, a formulation based on 
variables in which the conformal factor of the 3-metric and the
trace of the extrinsic curvature are factored out and evolved 
separately has been studied.  This conformal-traceless (CT)
formulation was first introduced
by Nakamura and Shibata~\cite{Shibata95} and later 
slightly modified by Baumgarte and Shapiro~\cite{Baumgarte99}.  
The stability properties of the CT formulation were shown
in \cite{Baumgarte99} to be better than those of the ADM formulation
for linear waves.  The improvement in numerical stability in the
CT formulation versus the ADM formulation was
demonstrated in strong field dynamical cases in~\cite{Alcubierre99d}.
A step toward understanding the improved stability properties of 
the CT formulation was taken in~\cite{Alcubierre99e} where it was
shown by analytically linearizing the ADM and CT equations about flat
space that the CT system effectively decouples the gauge modes and
constraint violating modes.  It was conjectured that giving the constraint
violating modes nonzero propagation speed results in a stable evolution.

Here, we take another step towards understanding the improved 
stability properties of the CT system by performing
a von Neumann stability analysis on discretizations of
both the ADM and CT systems.  We are led
to the von Neumann stability analysis by Lax's 
equivalence theorem~\cite{Lax56},
which states that given a well posed initial value problem and a
discretization that is consistent with that initial value problem
(i.e., the finite difference equations are faithful to the differential
equations), then stability is equivalent to convergence.  Here, the
words ``stability'' and ``convergence'' are taken to mean very
specific things.  Convergence is taken to mean pointwise convergence
of solutions of the finite difference equations to solutions of
the differential equations.  This is the {\it pi\`{e}ce de r\'{e}sistance}
of numerical relativity.  After all, what we are interested in are
solutions to the differential equations.  Stability, on the
other hand, has
a rather technical definition involving the uniform boundedness of
the discrete Fourier transform of the finite difference update
operator (see~\cite{Richtmyer67} for details).
In essence, stability is the statement that there should
be a limit to the extent to which any component of an initial discrete
function can be amplified during the numerical evolution procedure
(note that stability is a statement concerning the finite difference
equations, {\it not} the differential equations).  Fortunately, the 
technical definition of stability can be shown to be equivalent to
the von Neumann stability condition, which will be described in 
detail in the next section.  

While one cannot apply Lax's equivalence theorem directly in numerical
relativity (the initial value problem well-posedness 
assumption is not valid for the
Einstein field equations in that the evolution operator is not,
in general, uniformly bounded ), 
numerical relativists often use it as a ``road
map'';  clearly consistency and stability are important parts of
any discretization of the Einstein equations (curiously, convergence
is usually implicitly assumed in most
numerical studies).  Code tests, if done at all, usually
center around verifying the consistency of the finite difference equations
to the differential equations (as an example of the extents to
which some numerical relativists will go to check the consistency
of the finite difference equations to the differential equations, 
see, e.g.,~\cite{Font98b}).  Stability, on the other hand, is usually
assessed postmortem.  If the code crashes immediately after
a sharp rise in Nyquist frequency noise and/or if the code crashes
sooner in coordinate time at higher resolutions, the code is deemed unstable.  

We suggest that the stability of the code can be assessed before (and
perhaps even more importantly, while)
numerical evolutions take place.  As will be seen in the next section,
the stability properties of any given nonlinear finite difference update
operator depend not only on the Courant factor $\frac {\Delta t}{\Delta z}$,
but also on the values of the discrete evolution variables themselves.
Therefore, during numerical evolutions of nonlinear problems, as the 
evolved variables change from discrete timestep to timestep, the
stability properties of the finite difference operator change
along with them!  Ideally, one would want to verify that the 
finite difference update operator remains stable for {\it each} point
in the computational domain {\it at each timestep}.  While the
computational expense of this verification would be prohibitive,
verification at a reasonably sampled subset of discrete points
could be feasible.

The remainder of the paper is outlined as follows.  
Section~\ref{sec:vonneumann} will describe the von Neumann stability analysis
for a discretization of a general set of 
nonlinear partial differential equations
in one spatial dimension.  Included will be results from a von Neumann
stability analysis of the linear wave equation discretized with an 
iterative Crank-Nicholson scheme.  Section~\ref{sec:grflat} will present
the ADM and CT formulations of the Einstein equations restricted
to diagonal metrics and further restricted to dependence on only
one spatial variable.  The von Neumann stability analysis is 
performed on iterative Crank-Nicholson discretizations of both 
formulations with flat initial data.  Section~\ref{sec:nonlin_anal} will 
repeat the von Neumann stability analysis from section ~\ref{sec:grflat}
with a nonlinear wave solution for initial data.

%%%%%%%%%%%%%%%%%%%%%%%%%%%%%%%
%%%   VON NEUMANN ANALYSIS  %%%
%%%%%%%%%%%%%%%%%%%%%%%%%%%%%%%

\section{The von Neumann Stability Analysis}
\label{sec:vonneumann}

Let a set of partial differential equations be given as 
\begin{equation}
\label{eq:diffeq}
\frac {\partial \vec{u}}{\partial t} =
   S(\vec{u}, {\partial}_z \vec{u}, {{\partial}_z}^2 \vec{u})
\end{equation}
where $\vec{u} = \vec{u}(z,t)$ is a vector whose components consist of 
the dependent variables that are functions of the independent variables
$0 \leq z \leq N$ and $t \geq 0$ (while we ignore boundary conditions
in this paper, as we consider only interior points of equations with
finite propagation speeds, the von Neumann analysis presented here
is easily extended to include boundary treatments~\cite{Richtmyer67}).  

Now, consider a discretization of the independent variables $z$ and $t$:
\begin{eqnarray}
z_j = j \: \Delta z,  & \; \; j=0,1,2,...,m \\
t_n = n \: \Delta t,  & \; \; n=0,1,2,...
\end {eqnarray}
along with a discretization of the dependent variables $\vec{u}(z,t)$:
\begin{equation}
{\vec{u}}^n_j = \vec{u}(z_j, t_n).
\end{equation}
Furthermore, consider a consistent discretization of the set of differential
equations in the form of Eq.~\ref{eq:diffeq} that can be written in the form
\begin{equation}
\label{eq:finitediffeq}
{\vec{u}}^{n+1}_j = {\vec{S}}^n_j (...,{\vec{u}}^n_{j-2},
   {\vec{u}}^n_{j-1},{\vec{u}}^n_{j},{\vec{u}}^n_{j+1},
   {\vec{u}}^n_{j+2},...).
\end{equation}
As we will only be analyzing the iterative Crank-Nicholson 
discretization scheme, we assume a two-step method 
as written in Eq.~\ref{eq:finitediffeq}.  However, a von Neumann analysis
does not depend on this, and could easily be performed for three-step
methods such as the leapfrog scheme.

Now, assume some initial data 
\begin{equation}
{\vec{u}}^0_j, \; \; 0 \leq  j \leq m
\end{equation}
for the discrete variables is given at time $t=0$.
The amplification matrix at $z_j$ is given by~\cite{Richtmyer67}
\begin{equation}
\label{eq:ampmatrix}
{\bf G}^n_j = \sum_{j'=0}^m \frac {\partial {\vec{S}}^n_j}
   {\partial {\vec{u}}^n_{j'}}  \; e^{i \: \Delta\!z \: k_z \: (j - j')}
\end{equation}
The condition for numerical stability~\cite{Richtmyer67} is that the
spectral radius of the amplification matrix ${\bf G}^n_j$ be 1 or less
for each wavenumber $k_z$.
That is, each eigenvalue ${\lambda}_i$ 
of the amplification matrix ${\bf G}^n_j$ should
have a modulus of 1 or less for each discrete mode $k_z$
\begin{equation}
\left | {\lambda}_i \right | \leq 1.
\label{eq:vonneumann_condition}
\end{equation}

First, notice that for linear finite difference equations 
(Eq.~\ref{eq:finitediffeq}), the amplification matrix ${\bf G}^n_j$ does
{\it not} depend on either the initial data ${\vec{u}}^0_j$ nor on 
the spatial index $j$.  It only depends on the discretization 
parameters $\Delta z$ and $\Delta t$ as well as the mode
wavenumber $k_z$.   That is, for the linear case, once one
specifies the discretization parameters $\Delta z$ and $\Delta t$,
one only needs to verify that the eigenvalues of the 
amplification matrix ${\bf G}^n_j$ have a modulus of less than
or equal to 1 for all wavenumbers $k_z$ to insure the 
numerical stability of a numerical update operator,
regardless of initial data ${\vec{u}}^0_j$.  
Contrast this with the nonlinear case where the amplification
matrix ${\bf G}^n_j$ is not only a function of discretization parameters
$\Delta z$,  $\Delta t$ and the mode wavenumber $k_z$, 
but also of initial data ${\vec{u}}^0_j$.
In this case, not only must one verify 
that the amplification matrix ${\bf G}^n_j$
have a spectral radius of 1 or less for each spatial index $j$ 
(assuming the initial data ${\vec{u}}^0_j$ depends on $j$, which
it will in general), but one must carry out this verification at
{\it every} time step $n$, as the data ${\vec{u}}^n_j$ will, in 
general, change with increasing $n$.  So, in the nonlinear case, 
the amplification matrix ${\bf G}^n_j$ will depend on both the spatial
index $j$ and the temporal index $n$.  In principle, one
must verify that ${\bf G}^n_j$ have a spectral radius of 1 or less
for each mode wavenumber $k_z$, for each spatial index $j$
and for each time step $n$, in order to be confident that the
solutions of the finite difference equations are converging
to solutions of the differential equations (which is, after all,
what we are ultimately interested in).

%%%%%%%%%%%%%%%%%%%%%%%%%%%%%%%%
%%%   Linear Wave Equation   %%%
%%%%%%%%%%%%%%%%%%%%%%%%%%%%%%%%

\subsection{Example: Linear Wave Equation}
\label{sec:linearwaveeq}

\begin{figure}
\centerline{\psfig{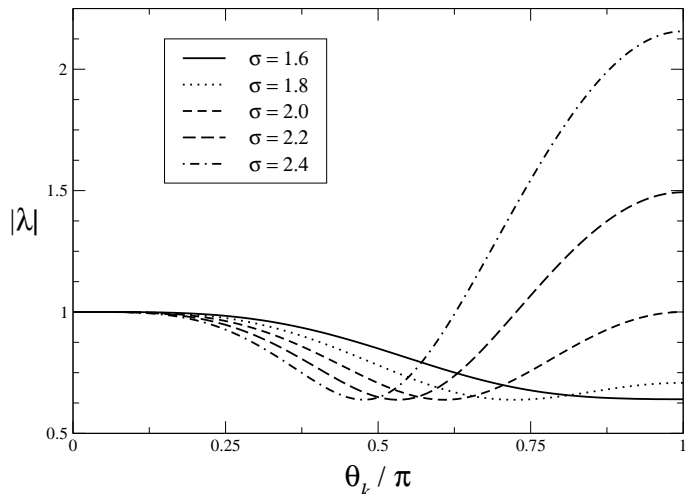}}
\caption{ The largest eigenvalue (ignoring eigenvalues
that are identically 1) of the amplification matrix
as a function of ${\theta}_k \equiv k_z \: \Delta z$
for the 2-iteration
iterative Crank-Nicholson discretization of the scalar wave equation.
Various values of 
the Courant factor $\sigma = \Delta t / \Delta z$ are shown.}
\label{fig:lineareigenvalue}
\end{figure}

In~\cite{Teukolsky00}, a von Neumann analysis of the advection equation
was presented for the iterative Crank-Nicholson scheme.  Here, as
a prelude to computing a von Neumann analysis for discretizations
of the equations
of general relativity, we present a von Neumann analysis for a 2-iteration
iterative Crank-Nicholson discretization of the wave equation in
1 dimension
\begin{equation}
\frac {{\partial}^2 \phi}{\partial t^2} = 
\frac {{\partial}^2 \phi}{\partial z^2}.
\end{equation}
We write the wave equation in first-order form as in Eq.~\ref{eq:diffeq}.
Defining $D \equiv \frac {\partial \phi}{\partial z}$, 
$\Pi \equiv \frac {\partial \phi}{\partial t}$, and
\begin{equation}
\vec{u} \equiv  \left [ \begin{array}{c} 
            \phi \\ D \\ \Pi
        \end{array} \right ],
\end{equation}
the wave equation in one spatial dimension becomes
\begin{equation}
\frac {\partial \vec{u}} {\partial t} =
\left [ \begin{array}{c}
            \Pi \vspace{0.2cm} \\ \vspace{0.2cm}
            \frac {\partial \Pi}  {\partial z} \\ 
            \frac {\partial D}  {\partial z} \end{array} \right ].
\end{equation}
The iterative Crank-Nicholson discretization procedure begins by taking
a FTCS (Forward Time Centered Space) step which is used to define the 
intermediate variables ${}^{(1)}{\tilde{\vec{u}}}^{n+1}_j$:
\begin{equation}
\frac {{}^{(1)}{\tilde{\vec{u}}}^{n+1}_j - \vec{u}^n_j}{\Delta t} = 
\left [ \begin{array}{c}
   {\Pi}^n_j \\
    ({ {\Pi}^n_{j+1} - {\Pi}^n_{j-1}}) / ( {2 \: \Delta z} ) \\
    ({ {D}^n_{j+1} - {D}^n_{j-1}} ) / ( {2 \: \Delta z})
        \end{array} \right ].
\end{equation}
This intermediate state variable is averaged with the original state
variable
\begin{equation}
{}^{(1)}{\vec{u}}^{n+1/2}_j = \frac {1}{2} (
{}^{(1)}{\tilde{\vec{u}}}^{n+1}_j + {\vec{u}}^n_j )
\end{equation}
which, in turn, is used to calculate ${}^{(1)}{\vec{u}}^{n+1}_j$, 
the state vector produced from the first iteration of the
iterative Crank-Nicholson scheme:
\begin{equation}
\frac {{}^{(1)}{\vec{u}}^{n+1}_j - \vec{u}^n_j}{\Delta t} = 
\left [ \begin{array}{c}
   {{}^{(1)}\Pi}^{n+1/2}_j \\
    ({ {{}^{(1)}\Pi}^{n+1/2}_{j+1} - 
           {{}^{(1)}\Pi}^{n+1/2}_{j-1}}) / ( {2 \: \Delta z}) \\
    ({ {{}^{(1)}D}^{n+1/2}_{j+1} - 
           {{}^{(1)}D}^{n+1/2}_{j-1}}) / ( {2 \: \Delta z} )
        \end{array} \right ].
\end{equation}
The averaged state ${}^{(2)}{\vec{u}}^{n+1/2}_j$ is calculated
\begin{equation}
{}^{(2)}{\vec{u}}^{n+1/2}_j = \frac {1}{2} (
{}^{(1)}{\vec{u}}^{n+1}_j + {\vec{u}}^n_j )
\end{equation}
and used to compute the final state variable ${}^{(2)}{\vec{u}}^{n+1}_j$
\begin{equation}
\frac {{}^{(2)}{\vec{u}}^{n+1}_j - \vec{u}^n_j}{\Delta t} =         
\left [ \begin{array}{c}
   {{}^{(2)}\Pi}^{n+1/2}_j \\
    ({ {{}^{(2)}\Pi}^{n+1/2}_{j+1} - 
           {{}^{(2)}\Pi}^{n+1/2}_{j-1}}) / ( {2 \: \Delta z} ) \\
    ({ {{}^{(2)}D}^{n+1/2}_{j+1} - 
           {{}^{(2)}D}^{n+1/2}_{j-1}}) / ( {2 \: \Delta z} )
        \end{array} \right ].
\end{equation}
Using this second iteration as the final iteration, we can write
Eq.~\ref{eq:finitediffeq} explicitly as the three equations
\begin{eqnarray}
   {\phi}^{n+1}_j & = & {\phi}^n_j + 
      \frac {\sigma \: \Delta t}{4} (D^n_{j+1} - D^n_{j-1}) + \nonumber \\
   & &   \frac { {\sigma}^2 \: \Delta t}{16} ({\Pi}^n_{j+2} - {\Pi}^n_{j-2}) +
      \Delta t (1 - \frac {{\sigma}^2}{16}) {\Pi}^n_j \\
   D^{n+1}_j & = & (1 - \frac {{\sigma}^2}{4}) D^n_j +
      \frac {{\sigma}^2}{8} (D^n_{j+2} + D^n_{j-2}) + \nonumber \\
   & & \frac {{\sigma}^3}{32} ({\Pi}^n_{j+3} - {\Pi}^n_{j-3}) +
      \frac {\sigma}{2} (1 - \frac{3}{16} {\sigma}^2) 
            ({\Pi}^n_{j+1} - {\Pi}^n_{j-1}) \\
   {\Pi}^{n+1}_j & = & (1 - \frac {{\sigma}^2}{4}) {\Pi}^n_j +
      \frac {{\sigma}^2}{8} ({\Pi}^n_{j+2} + {\Pi}^n_{j-2}) + \nonumber \\
   & & \frac {{\sigma}^3}{32} ({D}^n_{j+3} - {D}^n_{j-3}) +
      \frac {\sigma}{2} (1 - \frac{3}{16} {\sigma}^2) 
            ({D}^n_{j+1} - {D}^n_{j-1}),
\end{eqnarray}
where we have denoted $\sigma$ as the Courant factor $\Delta t / \Delta z$.
We compute the amplification matrix Eq.~\ref{eq:ampmatrix}, which
is independent of the spatial index $j$ and the time index $n$:
\begin{equation}
{\bf G} = \left [
   \begin{array}{ccc}
      G_{11} & G_{12} & G_{13} \\
      G_{21} & G_{22} & G_{23} \\
      G_{31} & G_{32} & G_{33} 
   \end{array}  \right ]
\end{equation}
where the components are
\begin{equation}
\begin{array}{l}
   G_{11} = 1 \nonumber \\
   G_{21} = G_{31} = 0 \nonumber \\
   G_{12} = - \frac {i \sigma \Delta t}{2} \sin{{\theta}_k} \\
   G_{13} = \Delta t (1 - \frac{{\sigma}^2}{8}) + 
              \frac {\Delta t {\sigma}^2}{8} \cos{(2 {\theta}_k)} \\
   G_{22} = G_{33} = (1 - \frac {{\sigma}^2}{4}) + 
                     \frac{{\sigma}^2}{4} \cos{(2 {\theta}_k)} \nonumber \\
   G_{23} = G_{32} = - \frac{i {\sigma}^3}{16} \sin{(3 {\theta}_k)} - 
              i \sigma ( 1 - \frac{3}{16} {\sigma}^2) \sin{{\theta}_k}, 
\end{array}
\end{equation}
where ${\theta}_k \equiv k_z \: \Delta z$.
The modulus of the eigenvalues of ${\bf G}$ are 
easily calculated and found to be
(ignoring the eigenvalue that is identically 1)
\begin{equation}
\label{eq:lineareigenvalue}
|{\lambda}_1| = |{\lambda}_2| = \sqrt{1 - 
   \frac {1}{4} {\sigma}^4 {\sin}^4 {\theta}_k +
   \frac {1}{16} {\sigma}^6 {\sin}^6 {\theta}_k}
\end{equation}
By inspection,
the von Neumann condition for stability, i.e. $|\lambda| \leq 1$
for all $k_z$, is
$\sigma \equiv \frac {\Delta t}{\Delta z} \leq 2$.  However,
it is instructive to look at the ${\theta}_k$ dependence of the stability
criterion.  Figure~\ref{fig:lineareigenvalue} shows
a plot of Eq.~\ref{eq:lineareigenvalue} as a function
of ${\theta}_k \equiv k_z \: \Delta z$ 
for various values of the Courant factor $\sigma$.  As can be
seen, the first eigenmode that goes unstable (i.e., has an eigenvalue whose
modulus is greater than 1) as $\sigma$ is increased is the mode
${\theta}_k \equiv k_z \: \Delta z = \pi$.  This corresponds to modes
of wavelength $2 \: \Delta z$, i.e. Nyquist frequency modes,
which are precisely the modes that
usually crop up and kill numerical relativity simulations.

\vspace{1.0cm}

%%%%%%%%%%%%%%%%%%%%%%%%%%%%%
%%%   General Relativity  %%%
%%%%%%%%%%%%%%%%%%%%%%%%%%%%%

\section{General Relativity}
\label{sec:grflat}
Here, we present the analytic equations for the ADM and CT formulations
that we will use to study the stability properties of numerical methods
in general relativity.   We will discretize the general relativity
evolution equations using the same discretization used
in the previous section for discretizing the scalar wave equation,
namely, a 2-iteration iterative Crank-Nicholson scheme.  Before analyzing
the stability of the discretization for non-linear waves, we will perform
a von Neumann analysis for the discretizations of both the ADM and CT
equations about flat space.

%%%%%%%%%%%%%%%%%%%%%%%%%
%%%   ADM Equations   %%%
%%%%%%%%%%%%%%%%%%%%%%%%%

\subsection{ADM Equations in 1-D (plane symmetry)}
\label{sec:admeq}

The form of the metric that we will use to study stability properties
of discretizations of the ADM form of the Einstein equations is given by
\begin{equation}
\label{eq:admmetric}
ds^2 = - {\alpha}^2 \: dt^2 \; + \; g_{xx} \: dx^2 \; + \;
   g_{yy} \: dy^2 \; + \; g_{zz} \: dz^2,
\end{equation}
where the lapse $\alpha$ and metric functions $g_{xx}$, $g_{yy}$, and 
$g_{zz}$ are functions of the independent variables $z$ and $t$.  
To put the evolution equations in first order form (Eq.~\ref{eq:diffeq}),
we introduce the extrinsic curvature functions $K_{xx}$, $K_{yy}$, 
and $K_{zz}$, each of which are also function of the independent variables
$z$ and $t$.  The evolution equations for this ADM system is given in
the form of Eq.~\ref{eq:diffeq} as
\begin{eqnarray}
\label{eq:adm}
\frac {\partial g_{ii}}{\partial t} & = & - 2 \alpha K_{ii} \nonumber \\
\frac {\partial K_{ii}}{\partial t} & = & \alpha R_{ii} -
   2 \alpha {K_i}^j K_{ij} + \alpha K K_{ii} - {\cal D}_i 
   {\cal D}_i \alpha \nonumber \\
\frac {\partial \alpha}{\partial t} & = & - \alpha K
\end{eqnarray}
where the index pair $ii$ takes on the values $\{xx,yy,zz\}$,
the index $j$ is summed over the values $\{x,y,z\}$, $K$ denotes the trace of
the extrinsic curvature $K = K_{xx}/g_{xx} + K_{yy}/g_{yy} + K_{zz}/g_{zz}$,
${\cal D}$ denotes the covariant derivative operator compatible with the 
3-metric, and $R_{ii}$ denotes the three components of the 
3-Ricci tensor, which are given explicitly as 
\begin{eqnarray}
R_{xx} & = & \frac {1}{4 g_{xx} g_{zz}} 
      {\left ( \frac{\partial g_{xx}}{\partial z} \right )}^2 - 
      \frac {1}{4 g_{yy} g_{zz}} \frac {\partial g_{xx}}{\partial z}
      \frac {\partial g_{yy}}{\partial z} + \nonumber \\
       &   &  \frac {1}{4 {g_{zz}}^2} \frac {\partial g_{xx}}{\partial z}
       \frac {\partial g_{zz}}{\partial z} - \frac {1}{2 g_{zz}} 
       \frac {{\partial}^2 g_{xx}}{\partial z^2} \nonumber \\
R_{yy} & = & \frac {1}{4 g_{yy} g_{zz}}
      {\left ( \frac{\partial g_{yy}}{\partial z} \right )}^2 -
      \frac {1}{4 g_{xx} g_{zz}} \frac {\partial g_{xx}}{\partial z}
      \frac {\partial g_{yy}}{\partial z} + \nonumber \\
       &   &  \frac {1}{4 {g_{zz}}^2} \frac {\partial g_{yy}}{\partial z}
       \frac {\partial g_{zz}}{\partial z} - \frac {1}{2 g_{zz}}
       \frac {{\partial}^2 g_{yy}}{\partial z^2} \\
R_{zz} & = & \frac {1}{4 {g_{xx}}^2} 
       {\left (\frac{\partial g_{xx}}{\partial z} \right)}^2 +
             \frac {1}{4 {g_{yy}}^2} 
       {\left (\frac{\partial g_{yy}}{\partial z} \right)}^2 + \nonumber \\
       &   & \frac {1}{4 g_{xx} g_{zz}} \frac {\partial g_{xx}}{\partial z}
      \frac {\partial g_{zz}}{\partial z} +
      \frac {1}{4 g_{yy} g_{zz}} \frac {\partial g_{yy}}{\partial z}
      \frac {\partial g_{zz}}{\partial z} - \nonumber \\
       &   & \frac {1}{2 g_{xx}}
       \frac {{\partial}^2 g_{xx}}{\partial z^2} - \frac {1}{2 g_{yy}}
       \frac {{\partial}^2 g_{yy}}{\partial z^2} \nonumber 
\end{eqnarray}
Throughout this paper, we use the so-called ``1 + log''
slicing condition in Eq.~\ref{eq:adm}.
This local condition on the lapse has been used successfully in several
recent applications~\cite{Miller99a,Alcubierre99b}.  
Moreover, it is a local condition, and thus, the 
von Neumann analysis remains local (this would be in contrast with
a global elliptic condition on the lapse, e.g. maximal slicing, in which
the calculation of the sum in the definition of the
amplification matrix in Eq.~\ref{eq:ampmatrix} would have nonzero global 
contributions).  

The initial value formulation is completed by specifying
initial data that satisfies the Hamiltonian and momentum constraints, given
respectively by
\begin{eqnarray}
{\cal H} & \equiv & {}^3 R + K^2 - K_{jk} K^{jk} = 0 \\
{\cal M}_z & \equiv & {\cal D}_j {K^j}_z - {\cal D}_z K
\end{eqnarray}

%%%%%%%%%%%%%%%%%%%%%%%%
%%%   CT Equations   %%%
%%%%%%%%%%%%%%%%%%%%%%%%

\subsection{Conformal-Traceless (CT) Equations in 1-D (plane symmetry)}
\label{sec:cteq}

The form of the metric that we will use to study stability properties
of the discretizations of the CT form of the Einstein equations,
as defined in~\cite{Baumgarte99,Alcubierre99d,Alcubierre99e} is given by
\begin{equation}
\label{eq:ctmetric}
ds^2 = - {\alpha}^2 \: dt^2 \; + \; e^{4 \phi}
   ({\tilde{g}}_{xx} \: dx^2 \; + \;
    {\tilde{g}}_{yy} \: dy^2 \; + \; {\tilde{g}}_{zz} \: dz^2),
\end{equation}
where the lapse $\alpha$, the conformal function $\phi$, and the
conformal metric components ${\tilde{g}}_{xx}$, ${\tilde{g}}_{yy}$,
and ${\tilde{g}}_{zz}$ are functions of the independent variables
$z$ and $t$.  The determinant of the conformal 3-metric
is identically 1.   Instead of evolving the extrinsic curvature components
as in the ADM formalism, the extrinsic curvature is split into
its trace ($K$) and traceless (${\tilde{A}}_{ii}$) components:
\begin{equation}
K_{ii} = e^{4 \phi} (\frac{1}{3} {\tilde{g}}_{ii} K  + {\tilde{A}}_{ii}).
\end{equation}
In addition, the conformal connection function ${\tilde{\Gamma}}^z$, defined
by
\begin{equation}
{\tilde{\Gamma}}^z = - {\partial}_j {\tilde{g}}^{jz}
\end{equation}
is also treated as an evolved variable.  There are therefore 10 evolution
equations that are in the form of Eq.~\ref{eq:diffeq}, and given explicitly
as 
\begin{eqnarray}
\label{eq:ct}
\frac {\partial \phi}{\partial t} & = & - \frac {1}{6} \alpha K \nonumber \\
\frac {\partial {\tilde{g}}_{ii}}{\partial t} & = & - 2 \alpha
   {\tilde{A}}_{ii} \nonumber \\
\frac{\partial K}{\partial t} & = & \alpha {\tilde{A}}_{jk} {\tilde{A}}^{jk} +
\frac {1}{3} \alpha K^2 - {\cal D}_j {\cal D}^j \alpha \nonumber \\
\frac {\partial {\tilde{A}}_{ii}}{\partial t} & = & \alpha 
   ( e^{-4 \phi} R_{ii} - \frac{1}{3} {\tilde{g}}_{ii} {\tilde{A}}^{jk}
      {\tilde{A}}^{jk} + \frac{2}{9} {\tilde{g}}_{ii} K^2 + 
        K {\tilde{A}}_{ij} - 2 {\tilde{A}}_{ij} {{\tilde{A}}_{i}}^j)- \\
        &  &  e^{-4 \phi} ({\cal D}_i {\cal D}_i \alpha - \frac{1}{3} g_{ii}
      {\cal D}_j {\cal D}^j \alpha) \nonumber \\
\frac {\partial {\tilde{\Gamma}}^{z}}{\partial t} & = & \alpha (
- \frac {4}{3 {\tilde{g}}_{zz}} \frac {\partial K}{\partial z} +
   \frac {12}{ {{\tilde{g}}_{zz}}^2} {\tilde{A}}_{zz} \frac {\partial \phi}
   {\partial z} + 2 {\tilde{\Gamma}^z}_{ij} {\tilde{A}}^{ij}) \nonumber \\
\frac {\partial \alpha}{\partial t} & = & - \alpha K \nonumber 
\end{eqnarray}
where the indices of ${\tilde{A}}_{ij}$ are raised and lowered with
the conformal metric ${\tilde{g}}_{ij}$, the index on the covariant
derivative operator ${\cal D}$ with respect to the physical metric
is raised and lowered with the physical metric $g_{ij} = e^{4 \phi}
{\tilde{g}}_{ij}$, and ${\tilde{\Gamma}^k}_{ij}$ are the 
Christoffel symbols related to the conformal metric ${\tilde{g}}_{ij}$.
Note that the Hamiltonian constraint has been substituted in for the
3-Ricci scalar in the equations for $\frac{\partial K}{\partial t}$ and
$\frac {\partial {\tilde{A}}_{ii}}{\partial t}$.  Also, the momentum constraint
has been substituted in the equation for 
$\frac {\partial {\tilde{\Gamma}}^{z}}{\partial t} $.  This corresponds to
the ``Mom'' system from~\cite{Alcubierre99d}, and the
$\sigma=1$, $m=1$, $\xi =0$ system from~\cite{Alcubierre99e}.  Note that the 
Ricci components $R_{ii}$ can be written in terms of the conformal Ricci
components ${\tilde{R}}_{ii}$:  
\begin{eqnarray}
R_{ii} & = & {\tilde{R}}_{ii} + 2 \frac {\partial \phi}{\partial z}
   {\tilde{\Gamma}^z}_{ii} - 2 {\tilde{g}}_{ii} ( {\tilde{\cal D}}_j 
   {\tilde{\cal D}}^j \phi + \frac {2}{{\tilde{g}}_{zz}} 
      {(\frac{\partial \phi}{\partial z})}^2 ) +  \\
 &  &     \delta_{iz} (4 {(\frac{\partial \phi}{\partial z})}^2 - 2
         \frac { {\partial}^2 \phi}{\partial z^2}),
\end{eqnarray}
where the indices of the covariant derivative operator $\tilde{\cal D}$ 
with respect to the conformal metric are raised and lowered
with the conformal metric.  The conformal Ricci components can
in turn be written as
\begin{eqnarray}
\tilde{R}_{ii} & = & - \frac {1}{2 \tilde{g}_{zz}} 
   \frac { {\partial}^2 \tilde{g}_{ii}}{\partial z^2} + 
   \delta_{iz} \tilde{g}_{zz} \frac {\partial \tilde{\Gamma}^z}{\partial z} \\
  &  &  + \frac {1}{2} \tilde{\Gamma}^z \frac {\partial \tilde{g}_{ii}}
         {\partial z} + {\tilde{\Gamma}^j}_{zk} 
      ( 2 {{\tilde{\Gamma}_z}{}^k}_j + {{\tilde{\Gamma}_j}{}^k}_z)
\end{eqnarray}
Notice that, although we have imposed planar symmetry, we have
expressed the Ricci tensor in terms of derivatives of the 
conformal metric $\tilde{g}_{ij}$ and of the conformal connection
function $\tilde{\Gamma}^z$ just as is done for full 3-D numerical
relativity~\cite{Baumgarte99,Alcubierre99d}.

%%%%%%%%%%%%%%%%%%%%%%%%%%%%%%%%%%%%%%%%%%%
%%%   FLAT SPACE VON NEUMANN ANALYSIS   %%%
%%%%%%%%%%%%%%%%%%%%%%%%%%%%%%%%%%%%%%%%%%%

\subsection{von Neumann stability analysis about flat space}
\label{sec:stab_flat}

As outlined for the scalar wave equation in Section~\ref{sec:linearwaveeq}, 
we discretize both the ADM equations (Eq.~\ref{eq:adm}) and the
CT equations (Eq.~\ref{eq:ct}) using a 2-iteration
iterative Crank-Nicholson scheme.  The resulting finite
difference equations, even though
planar symmetry and a simplified form of the metric is assumed, are still
too complicated to perform a von Neumann analysis by hand.  The complication
arises due to the recursive nature of the iterative Crank-Nicholson
method;  a 2-iteration procedure results in the source terms of 
Eqs.~\ref{eq:adm}~and~\ref{eq:ct} being computed 3 times in 
recursive succession.  

We have performed the von Neumann analysis of the ADM and CT equations
in two independent ways.  In the first way, we
used the symbolical calculation computer program MATHEMATICA to explicitly
calculate the 2-iteration iterative Crank-Nicholson update, and then
explicitly calculated the amplification matrix Eq.~\ref{eq:ampmatrix} in
terms of the initial data variables.  We then took these expressions,
substituted the initial data about which we want to compute the von Neumann
analysis, and calculated the eigenvalues to arbitrary precision inside
MATHEMATICA.   The second, independent way of performing the 
von Neumann analysis was to write an evolution code for the 
finite difference equations.  We then input the initial
data about which we want to compute the von Neumann analysis
and computed the derivatives in the definition of the
amplitude matrix Eq.~\ref{eq:ampmatrix} using finite differencing
(this amounts to finite differencing the finite difference equations!).
The package EISPACK was then used to compute the eigenvalues
of the resulting amplification matrices.  To obtain the highest accuracy
possible, whenever calculating the amplification matrix Eq.~\ref{eq:ampmatrix}
using finite differencing, we finite difference with multiple 
discretization parameters $h$, and use Richardson extrapolation
to obtain values of the derivatives.  

Both methods were used, and found to produce identical results.  All results
reported in the paper were produced using both methods as described
above.  We emphasize the use of two independent methods not only
to verify our results, but also for more practical reasons:  one may
eventually want to perform a von Neumann analysis for a full 3-D numerical
relativity code.  The analytic method using a symbolical manipulation package
such as MATHEMATICA may not be feasible in the near future.  It took
well over 100 hours on one node of an Origin 2000  running MATHEMATICA
to perform the symbolical calculations needed for the von Neumann analysis
of a plane symmetric code.  It may be several orders of magnitude more
expensive to analyze a full 3D evolution update.
The finite difference method for
computing the amplification matrix is much quicker, and it is 
reassuring that the finite difference method, used in conjunction with
Richardson extrapolation, is accurate enough to reproduce the same 
detailed structures of the eigenvalues of the amplification matrices
as doing the analytic calculation.

\vspace{1.0cm}

\begin{figure}
\centerline{\psfig{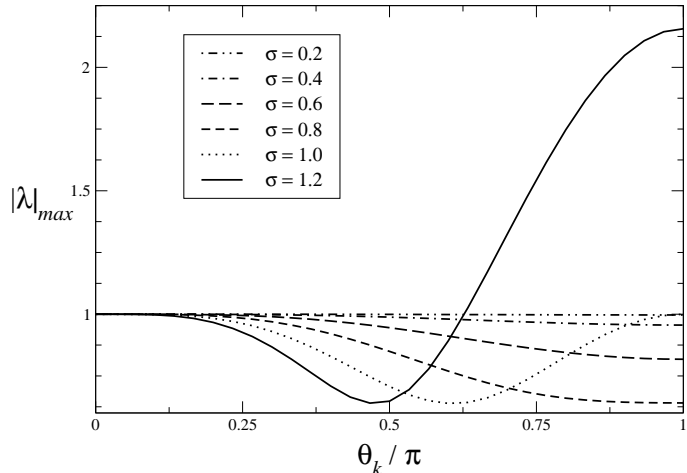}}
\caption{ The largest eigenvalue (ignoring eigenvalues
that are identically 1) of the amplification matrix as
a function of ${\theta}_k \equiv k_z \: \Delta z$ for the 2-iteration
iterative Crank-Nicholson discretization of the ADM equations with
flat initial data.
Various values of the Courant factor
$\sigma = \Delta t / \Delta x$ are shown.}
\label{fig:adm_flat}
\end{figure}

In Figure~\ref{fig:adm_flat} we plot the maximum of the modulus
of the eigenvalues (neglecting eigenvalues that are exactly 1) of
the amplification matrix of
the 2-iteration iterative Crank-Nicholson discretization scheme
of the ADM equations from Section~\ref{sec:admeq} using flat
space as initial data, namely, $g_{ii} = \alpha = 1$, and 
$K_{ii} = 0$.  We see that 
the spectral radius of the amplification matrix is less than
or equal to 1 for $\sigma = \Delta t / \Delta x \leq 1$.  
Notice that for the Nyquist frequency mode 
($\theta_k \equiv k_z \: \Delta z = \pi$), when the Courant
factor $\sigma$ is $1$, all eigenvalues of the amplification matrix
have a modulus of exactly 1.
For $\sigma > 1$,  all Nyquist frequency modes are unstable
(i.e. they all have amplification matrices with spectral radii
$> 1$).

\begin{figure}
\centerline{\psfig{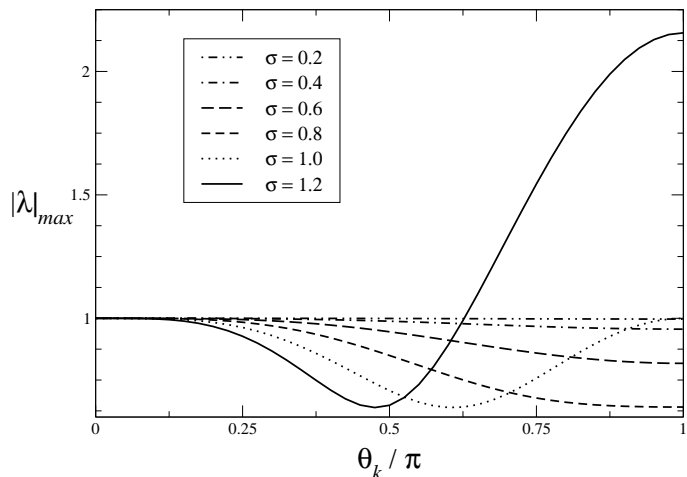}}
\caption{ The largest eigenvalue (ignoring eigenvalues
that are identically 1) of the amplification matrix as
a function of ${\theta}_k \equiv k_z \: \Delta z$ for the 2-iteration
iterative Crank-Nicholson discretization of the CT equations with
flat initial data.
Various values of the Courant factor
$\sigma = \Delta t / \Delta x$ are shown.}
\label{fig:ct_flat}
\end{figure}

In Figure~\ref{fig:ct_flat} we plot the maximum of the modulus
of the eigenvalues (neglecting eigenvalues that are exactly 1) of
the amplification matrix of the 2-iteration iterative Crank-Nicholson 
discretization scheme of the CT equations from Section~\ref{sec:cteq}, again
using flat space as initial data.  The resulting plot is identical to
that of the ADM equations for flat space.  The stability properties
of the CT equations about flat space 
are exactly the same as the stability properties
of the ADM equations about flat space.

\vspace{1.0cm}

%%%%%%%%%%%%%%%%%%%%%%%%%%%%%%%%%%%%%%%%%%%%%%%%%%%%%%%%
%%%   Von Neumann analysis for nonlinear planewaves  %%%
%%%%%%%%%%%%%%%%%%%%%%%%%%%%%%%%%%%%%%%%%%%%%%%%%%%%%%%%

\section{Von Neumann analysis for nonlinear plane waves}
\label{sec:nonlin_anal}

\begin{figure}
\centerline{\psfig{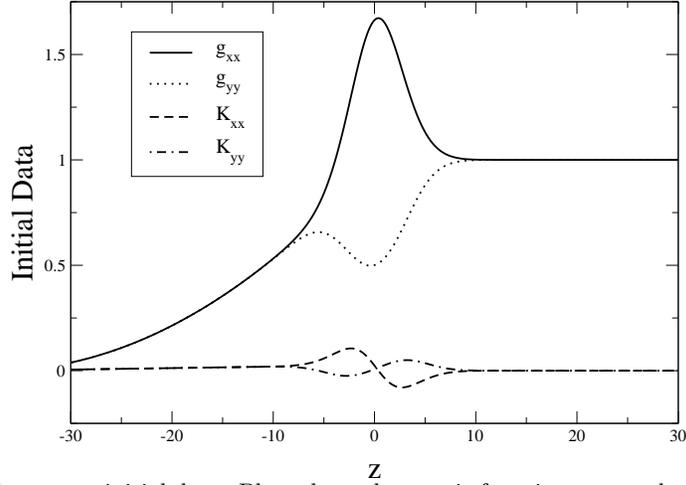}}
\caption{ Nonlinear plane wave initial data.  Plotted are the metric
functions $g_{xx}$ and $g_{yy}$ as well as the extrinsic
curvature functions $K_{xx}$ and $K_{yy}$. The metric function
$g_{zz}$ as well as the lapse $\alpha$ are identically 1, and the
extrinsic curvature component $K_{zz}$ is identically 0. }
\label{fig:mtwdata}
\end{figure}

In this section, we study the stability properties of the ADM
and CT equations about nonlinear plane waves.  We require
initial data that corresponds to nonlinear plane waves that satisfy 
the constraints and takes on the form of our simplified
metric Eqs.~\ref{eq:admmetric}~and~\ref{eq:ctmetric}.  
We choose an exact plane wave solution
first given by~\cite{Bondi59}.  The metric
is assumed to take the form
\begin{equation}
ds^2 = - dt^2 \; + \; L^2(e^{2 \beta} \: dx^2 \; + \;
   e^{-2 \beta} \: dy^2) \; + \;  dz^2,
\label{eq:nonlin_metric}
\end{equation}
where $L = L(v)$, $\beta = \beta(v)$, and $v = t-z$.  Given an arbitrary
function $\beta(v)$, the Einstein equations reduce to the 
following ordinary differential equation for L(v):
\begin{equation}
\label{eq:einstein_plane}
\frac {d^2 L(v)}{dv^2} + {\left ( \frac {d \beta (v)}{dv} \right )}^2 
   L(v) = 0
\end{equation}
In this paper, we take $\beta(v)$ to be given as
\begin{equation}
\beta (v) = \frac {3}{10} e^{- {v^2}/{16}}
\label{eq:nonlin_beta}
\end{equation}
Eq.~\ref{eq:einstein_plane} is then solved with a 4th order Runge-Kutta
solver.  Initial data is obtained by setting $t=0$ (and thus, $v = -z$),
shown in Figure~\ref{fig:mtwdata}.  We present results of a von Neumann
analysis about the point $z=3.0$ with $\Delta z = 0.15$.  We have 
investigated different values of $z$ and $\Delta z$, and find the 
results presented to be generic for different values of $z$
and $\Delta z$, as long as we remain in the nonlinear
regime, $z < 8$.

\vspace{1.0cm}

\begin{figure}
\centerline{\psfig{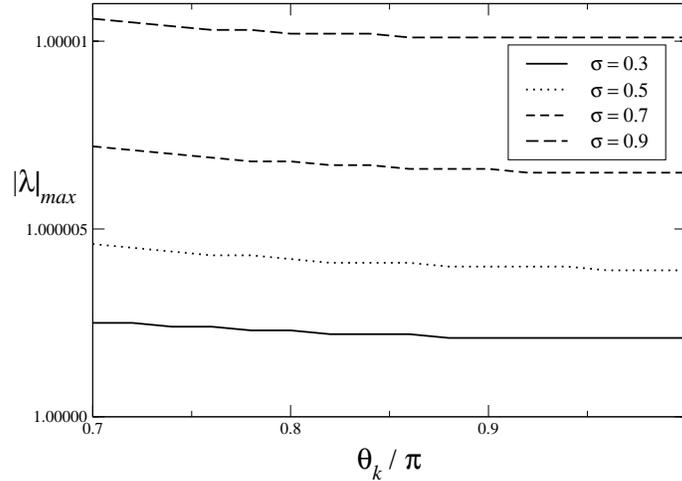}}
\caption{ The largest eigenvalue (ignoring eigenvalues that
are exactly 1) of the amplification matrix as a function of 
${\theta}_k \equiv k_z \: \Delta z$
for the Crank-Nicholson discretization of the 
ADM system about the nonlinear plane wave 
initial data at $z=0.3$ with $\Delta z = 0.15$.  Different values
of the Courant factor $\sigma = \Delta t / \Delta z$ are shown. }
\label{fig:admmedsig}
\end{figure}

\begin{figure}
\centerline{\psfig{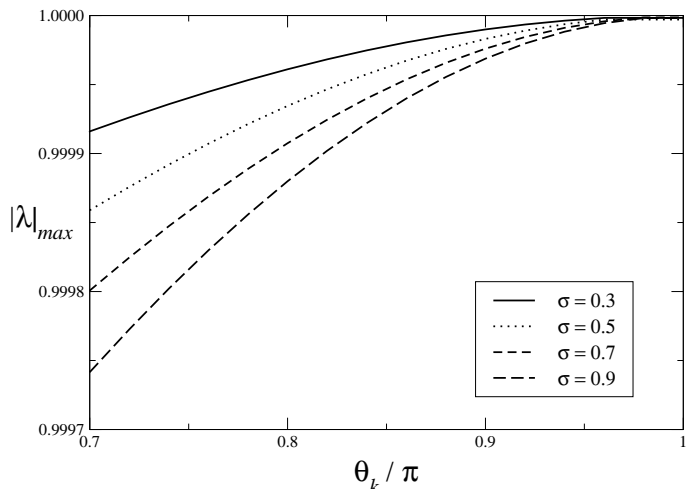}}
\caption{ The largest eigenvalue (ignoring eigenvalues that
are exactly 1) of the amplification matrix as a function of
${\theta}_k \equiv k_z \: \Delta z$
for the Crank-Nicholson discretization of the
CT system about the nonlinear plane wave
initial data at $z=0.3$ with $\Delta z = 0.15$.  Different values
of the Courant factor $\sigma = \Delta t / \Delta z$ are shown. }
\label{fig:ctmedsig}
\end{figure}

The main difference between the stability properties of discretizations
of the CT and ADM systems is shown in 
Figures~\ref{fig:admmedsig}~and~\ref{fig:ctmedsig}.  These show,
respectively, plots of the maximum modulus of the eigenvalues (ignoring
eigenvalues that are exactly 1) of the amplification matrices for 
discretizations of the ADM and CT systems at $z=3.0$.  These plots show
the range of the Courant factor 
$0.3 \leq \sigma \equiv \Delta t / \Delta z \leq 0.9$ and the range of modes
$0.7 \pi \leq {\theta}_k \equiv k_z \: \Delta z  \leq \pi$.  Notice how, for
these ranges of Courant factor $\sigma$ and mode wavenumber $k_z$, all of the
eigenvalues for the CT system are less than or equal to 1, where as for
the ADM system, there is always at least one eigenvalue that has a modulus
that is greater than 1.  For these ranges of $\sigma$ and $k_z$, we conclude
that the CT system is stable, while the ADM system is unstable.  

\vspace{1.0cm}

\vspace{1.0cm}

\begin{figure}
\centerline{\psfig{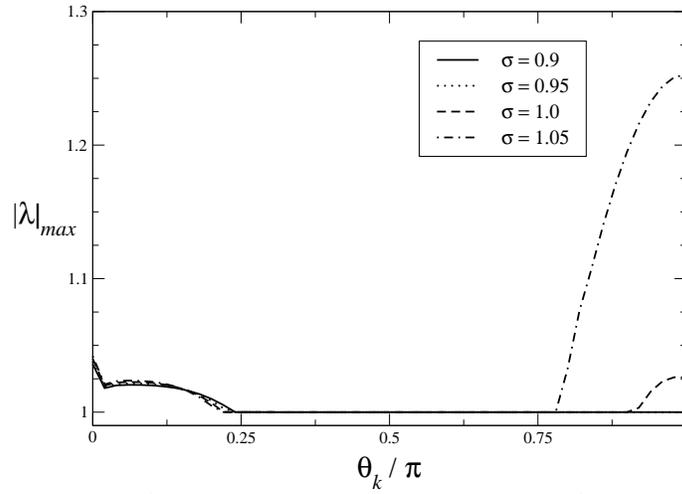}}
\caption{ The largest eigenvalue (ignoring eigenvalues that
are exactly 1) of the amplification matrix as a function of
${\theta}_k \equiv k_z \: \Delta z$
for the Crank-Nicholson discretization of the
ADM system about the nonlinear plane wave
initial data at $z=0.3$ with $\Delta z = 0.15$.  Values
for the Courant factor $ 0.9 \leq \sigma \leq 1.05$ are shown. }
\label{fig:admhighsig}
\end{figure}

\begin{figure}
\centerline{\psfig{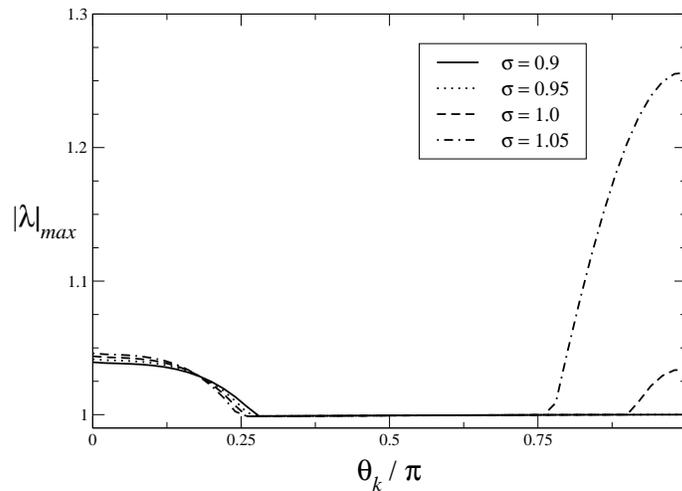}}
\caption{ The largest eigenvalue (ignoring eigenvalues that
are exactly 1) of the amplification matrix as a function of
${\theta}_k \equiv k_z \: \Delta z$
for the Crank-Nicholson discretization of the
CT system about the nonlinear plane wave
initial data at $z=0.3$ with $\Delta z = 0.15$.  Values
for the Courant factor $ 0.9 \leq \sigma \leq 1.05$ are shown. }
\label{fig:cthighsig}
\end{figure}

\begin{figure}
\centerline{\psfig{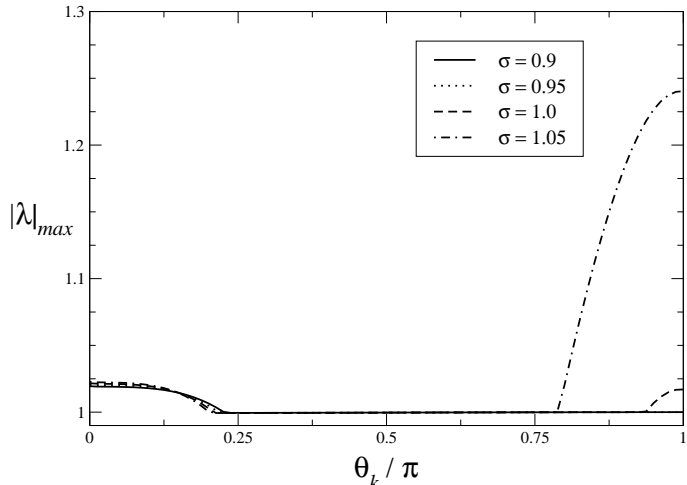}}
\caption{ The largest eigenvalue (ignoring eigenvalues that
are exactly 1) of the amplification matrix as a function of
${\theta}_k \equiv k_z \: \Delta z$
for the Crank-Nicholson discretization of the
CT system about the nonlinear plane wave
initial data at $z=0.3$ with $\Delta z = 0.075$.  Values
for the Courant factor $ 0.9 \leq \sigma \leq 1.05$ are shown. }
\label{fig:cthighsig2}
\end{figure}

By looking at Courant factors of values $\sigma > 1$, we see the typical
Nyquist frequency instability, shown in
Figures~\ref{fig:admhighsig}~and~\ref{fig:cthighsig} for the
ADM system and CT system, respectively.
Notice in both Figures~\ref{fig:admhighsig}~and~\ref{fig:cthighsig}
that the largest modulus of the eigenvalues of the amplification
matrices for long wavelength modes 
(${\theta}_k \equiv k_z \: \Delta z < 0.3 \pi $)
are all greater than $1$.  In fact, this is the
case for all values of 
Courant factor $\sigma$.  This is due to the fact that there
is an exponentially growing gauge mode in the analytic solution
to the analytic equations.  Recall that we are using a different
gauge choice than that given by the exact solution 
in Eqs.~\ref{eq:nonlin_metric}~-~\ref{eq:nonlin_beta}.  Using
the gauge choice given in Eq.~\ref{eq:adm} and
Eq.~\ref{eq:ct}, there exists an exponentially growing gauge
mode in $K_{zz}$.  To take into account equations that admit 
exponentially growing solutions, the von Neumann condition,
Eq.~\ref{eq:vonneumann_condition}, must be modified (see~\cite{Richtmyer67}
for details).  In order that finite difference discretizations
of equations that admit solutions that have exponentially growing modes
remain stable, we must have
\begin{equation}
\left | {\lambda}_i \right | \leq 1 + {\cal{O}}(\Delta t).
\label{eq:vonneumann_condition2}
\end{equation}
To verify that the long wavelength 
(${\theta}_k \equiv k_z \: \Delta z < 0.3 \pi$)
phenomena observed in Figures~\ref{fig:admhighsig}~and~\ref{fig:cthighsig}
is simply due to the existence of (long wavelength) 
exponentially growing modes in the analytic solution to the analytic 
equations, we repeat the calculations leading to 
Figures~\ref{fig:admhighsig}~and~\ref{fig:cthighsig}, but decrease
the discretization parameter $\Delta z$ by a 
factor of 2 ($\Delta z = 0.15/2 = 0.075$).  The results for both
the ADM and CT systems are similar, so we only show the results
for the CT system in Figure~\ref{fig:cthighsig2}.  Notice that 
by decreasing $\Delta z$ by a factor of 2 and holding the 
Courant factor $\sigma$ constant, we also decrease $\Delta t$ by
a factor of 2.  By comparing the long wavelength 
(${\theta}_k \equiv k_z \: \Delta z < 0.3 \pi$) sections of 
Figures~\ref{fig:cthighsig}~and~\ref{fig:cthighsig2}, we indeed see that
the difference between the maximum modulus of the eigenvalues of the
amplification matrices and 1 decreases by a factor of 2 when
$\Delta t$ is decreased by a factor of 2.  Therefore, the
discretizations of both the 
ADM and CT systems are stable for the long wavelength, exponentially
growing gauge mode.  Of course, the maximum eigenvalues 
for the high frequency sections of
Figures~\ref{fig:cthighsig}~and~\ref{fig:cthighsig2} 
for $\sigma > 1$ do not approach 1 as ${\cal{O}}(\Delta t)$, 
which signifies a true von Neumann instability for
$\sigma > 1$.

\vspace{1.0cm}

\vspace{1.0cm}

\vspace{1.0cm}

%%%%%%%%%%%%%%%%%%%%%%%
%%%   CONCLUSIONS   %%%
%%%%%%%%%%%%%%%%%%%%%%%

\section{Discussion and Conclusions}
\label{sec:conclusions}

First, while the results presented in
this paper are specific to the discretization
method and initial data chosen here, one of the main
points of this paper is to show that a von Neumann analysis can indeed be
applied directly to discretizations of the Einstein equations.
In the past, particular discretization methods were only analyzed 
with the von Neumann method through simple equations, such as the
linear wave equation.  Here we show that it is possible to carry
out a von Neumann analysis on discretizations of equations as
complicated as the Einstein equations.  We would like to
point out that a von Neumann analysis could also be
used to test
and/or formulate boundary
conditions.  The stability of outer boundary conditions,
as well as the stability of inner boundary
conditions for black hole evolutions, could be tested first
through a von Neumann analysis instead of the traditional (and painful) method
of coding up an implementation and looking for (and usually finding)
numerical instabilities during the numerical evolution.

In this paper, we have shown that the stability 
properties, as determined by a von Neumann
stability analysis, of a common discretization (a 2-iteration iterative
Crank-Nicholson scheme) of the ADM and CT systems about flat space
are similar to the stability properties of the scalar wave equation.  
However, as we would like to emphasize again, the stability properties
of a nonlinear finite difference update operator depend on the values of the
discrete evolution variables.  Therefore, it is not enough to study
the stability of numerical relativity codes about flat space.  In
principle, one must verify the stability of a nonlinear
finite difference update operator
about every discrete state encountered during the entire discrete evolution 
process, as argued in Section~\ref{sec:vonneumann}.   As a first
step in this direction, we studied the stability properties
of the ADM and CT systems about highly nonlinear plane waves by 
performing a von Neumann analysis for these scenarios.  Several 
interesting features presented themselves.  

The main difference between the stability of the
ADM and CT systems about the nonlinear wave solution is seen
in Figures~\ref{fig:admmedsig}~and~\ref{fig:ctmedsig}.  There, we
see that, for a wide range of Courant factors
$\sigma = \Delta t / \Delta z$ and wavenumbers $k_z$ (which includes the
mode that is usually the most troublesome in 
numerical relativity, the Nyquist frequency
mode whose wavelength is $2 \: \Delta z$), the CT system is stable 
(i.e., the amplification matrix has a spectral radius of 1 or less)
whereas the ADM system is unstable (i.e., the amplification
matrix has a spectral radius greater than 1).  Note that the ADM
system, in this region of Courant factor $\sigma$ and wavenumber $k_z$,
has an amplification matrix whose spectral radius is approximately 
$1 + 10^{-5}$.  This is a very small departure from unity.  Thus,
instabilities arising from these modes could take a long time to
develop during numerical evolutions.  For example, the 
2-iteration iterative Crank-Nicholson update operator for the 
scalar wave equation from Section~\ref{sec:linearwaveeq} has
a spectral radius of $1 + 10^{-5}$ for a Courant factor
$\sigma = \Delta t / \Delta z = 2.000004 $, and this
update operator can be used to evolve initial data sets many 
wavelengths before the Nyquist frequency instability sets in.  

Again, it must be pointed out that these results are specific to
plane wave spacetimes, simplified (diagonal) forms of the metric, 
choice of gauge, choice of initial data used here, and
the 2-iteration iterative Crank-Nicholson scheme.  It remains to be seen
whether or not these results are generic to other, more general 
discretizations of the Einstein equations and for more general data,
such as black hole and neutron star discrete evolutions.  The one
thing that is certain is that the von Neumann analysis can be used
as a diagnostic tool for determining the stability of discretizations
of nonlinear differential equations as complicated as the Einstein
equations.

%%%%%%%%%%%%%%%%%%%%%%%%%%%%
%%%   ACKNOWLEDGEMENTS   %%%
%%%%%%%%%%%%%%%%%%%%%%%%%%%%

\acknowledgements We would like to thank Josh Goldberg, Alyssa Miller,
David Rideout, Peter Saulson, Rafael Sorkin, 
and Wai-Mo Suen for encouragement and
interesting discussions.  This research
is supported by NSF (PHY 96-00507 and PHY 99-79985) and NASA (NCCS5-153).

%%%%%%%%%%%%%%%%%%%%%%
%%%   REFERENCES   %%%
%%%%%%%%%%%%%%%%%%%%%%

\bibliographystyle{prsty}

%\bibliography{bibtex/references}

\end{document}